\documentclass[showpacs,aps,prb,twocolumn]{revtex4}
\usepackage{amsmath}
\usepackage[final,dvips]{graphicx}
\usepackage{amsfonts}
\usepackage{amssymb}
\begin{document}
\title{Edge critical currents of dense Josephson vortex lattice in layered superconductors}
\author{A.\ E.\ Koshelev}
\affiliation{Materials Science Division, Argonne National Laboratory,Argonne, Illinois 60439}
\date{\today}

\begin{abstract}
We calculate the field dependence of the critical current for the
dense Josephson vortex lattice which is created by large magnetic
field $B$ applied along the layers of atomically layered
superconductors.
In clean samples a finite critical current appears due to the
interaction of the lattice with the boundaries. The boundary
induces an alternating deformation of the lattice decaying inside
the sample at the typical length, which is larger than the
Josephson length and increases proportional to the magnetic field.
The exact shape of this deformation and the total current flowing
along the surface are uniquely determined by the position of the
lattice in the bulk. The total maximum Josephson current has
overall $1/B$ dependence with strong oscillations. In contrast to
the well-known Fraunhofer dependence, the period of oscillations
corresponds to adding \emph{one flux quantum per two junctions}.
Due to interaction with the boundaries, the flux-flow voltage for
slow lattice motion also oscillates with field, in agreement with
recent experiments.
\end{abstract}
\maketitle

\section{Introduction}

Atomically layered superconductors, such as
Bi$_{2}$Sr$_{2}$CaCu$_{2}$O$_{x}$ (BSCCO), behave as stacks of
Josephson junctions. This so-called intrinsic Josephson effect has
been a subject of intense research in the past decade (see reviews
\cite{IJJReviews,YurgensReview}). In spite of atomic-size
separation between the neighboring junctions, in zero magnetic
field the system approximately behaves like  an array of
independent junctions. A weak dynamic interaction between the
neighboring junctions appears due to nonequilibrium effects
\cite{nonequil}.

\begin{figure}
[ptb]
\includegraphics[width=0.5\textwidth ] {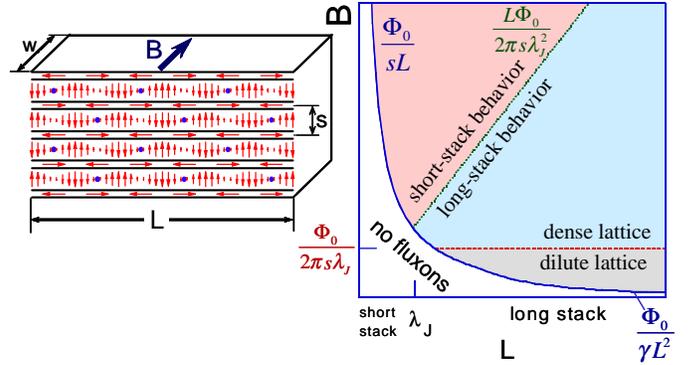}
\caption{\emph{Left part:} Dense Josephson vortex lattice in a
junction stack. Arrows show schematically distribution of the
Josephson and in-plane currents. Points mark centers of the
Josephson vortices. \emph{Right part:} Schematic field-size
diagram of a Josephson junction stack. Length and field scales are
set by the Josephson length $\lambda_J$ and the period of layered
structure $s$. This paper is focused on the long-stack/dense
lattice regime in the upper right part of the diagram.}
\label{Fig-DenseLatPhDiag}
\end{figure}

Situation is very different in the magnetic field applied along
the layer direction, which generates the Josephson vortex lattice.
Recently, dynamic properties of this lattice in BSCCO have been
subject of extensive experimental research
\cite{Lee,Hechtfischer,Latyshev,Ooi01}.
At high fields (above 0.5 tesla for BSCCO) the Josephson vortices
homogeneously fill all layers \cite{BulClemPRB91,KorshLarkPRB92}
(dense Josephson vortex lattice, see Fig.\
\ref{Fig-DenseLatPhDiag}). Josephson vortex arrays in neighboring
layers have strong inductive interaction \cite{Inductive}, which
strongly influences static and dynamic properties of the lattice.
Due to this interaction the magnetic properties of the stack are
very different from the magnetic properties of an isolated
junction.
A transport current flowing across the layers exerts a Lorentz
force on the Josephson vortices along the layer direction. An
important parameter is the \emph{critical current} above which the
lattice starts to move producing a finite voltage. This current is
determined either by bulk pinning or by interaction with the
boundaries. In this paper we consider the case of a homogeneous
junctions and neglect bulk pinning. The simplest and most known
case is a single small Josephson junction without inhomogeneities,
where the field dependence of the critical current is given by the
Fraunhofer dependence, $I_{c}(\Phi
)=I_{c0}|\sin(\pi\Phi/\Phi_{0})|/(\pi\Phi/\Phi_{0})$, with $\Phi$
being the magnetic flux through the junction. Observation of this
dependence has been considered as an important confirmation of the
{\it dc} Josephson effect. At present, the Fraunhofer field
dependence is routinely used as an indicator of the junction
homogeneity. The same dependence is also expected for the junction
stack with the lateral size smaller than the Josephson length
\cite{BulPRB92}. In a single long junction the field dependence of
the critical current has rather complicated structure due to
multiple coexistent states of the lattice \cite{LongJosJunct}.

In this paper we consider the critical current and lattice
structure near boundaries at high magnetic fields in the regime of
dense lattice in the stack of Josephson junctions. In this regime
the Josephson coupling can be treated as a small perturbation,
which allows to develop a full analytical description of this
system. Another important simplification is that at high fields
one can neglect the current self-field and neglect the difference
between the magnetic field inside the junction and the external
applied magnetic field.

We find that the boundary induces an alternating deformation of
the lattice. Averaging out the rapid phase oscillations, we derive
that the lattice deformation obeys the Sine-Gordon equation. The
deformation decays inside superconductor at the typical length
$l_B$, which is larger than the Josephson length and increases
proportional to the magnetic field, $l_B\propto B$. Structure of
the surface deformation and the total current flowing along the
surface is uniquely determined by the lattice position far away
from the boundary. The total current flowing through the stack is
given by the sum of two independent surface currents flowing at
the edges of the sample. Due to the commensurability effects, the
maximum current through the stack has oscillating field
dependence, which resembles the Fraunhofer dependence: it has
strong oscillations and overall $1/B$ dependence. However, the
period of these oscillations corresponds to adding \emph{one flux
quantum per two junctions}.

We also calculate the flux-flow voltage for the Josephson vortex
lattice slowly moving through a finite stack. Due to the
interaction with the boundaries this voltage also has an
oscillating contribution with period one flux quantum per two
junctions. Recently, such oscillations have been observed in BSCCO
mesas \cite{Ooi01}, and also were seen in numerical simulations
\cite{Machida02}. We find that the oscillating voltage normalized
to the bare flux-flow voltage has universal dependence on the
magnetic flux per one junction and the current density normalized
to the Josephson current density. At high currents the relative
amplitude of voltage oscillations decreases with current and field
as $1/J^2H^2$. We also show that in the case of dominating
in-plane dissipation, typical for BSCCO, the \emph{absolute}
amplitude of voltage oscillations weakly depends on magnetic field
in a wide field range, in agreement with experiment \cite{Ooi01}.

The new length scale $l_B$ increasing with the magnetic field also
sets the new field scale $B_L$, at which this length becomes of
the order of the junction length $L$. Above this field the system
crosses over to the regime of small junction and the field
dependence of the critical current crosses over to the usual
Fraunhofer dependence. Regimes of different behavior for a finite
stack in magnetic field are summarized in the phase diagram shown
in the right part of Fig.\ \ref{Fig-DenseLatPhDiag}.

The paper is organized as follows. In Section \ref{Sec:GenRel} we
write general relations describing the phase distribution for
layered superconductor in the high magnetic field applied along
the layers. In Section \ref{Sec:LatStruc} we consider lattice
structure near the surface, calculate the surface energy and
surface current. In Section \ref{Sec:CritCur} we calculate the
oscillating field dependence of the critical current in a finite
stack. In Section \ref{Sec:Voltage} we consider the field
oscillations of the flux-flow voltage due to the interaction with
the boundaries for slow lattice motion.

\section{\label{Sec:GenRel} General relations}

Consider a stack of the Josephson junctions with length
$L\gg\lambda_{J}\equiv\gamma s$, where $\gamma$ is the anisotropy
parameter, in high magnetic field $B\gg B_{cr}=\Phi_{0}/2\pi\gamma
s^{2}$ applied along the direction of the layers (see Fig.\
\ref{Fig-DenseLatPhDiag}). At high field the screening effects can
be neglected and the phase distribution $\varphi_{n}(y)$ is
described by the energy (per junction)
\begin{widetext}
\begin{equation}
\mathcal{E}_{J}=\frac{w}{N}\sum_{n}\int_{0}^{L}dy\left[
\frac{J}{2}\left( \frac{d\varphi_{n}}{dy}\right)
^{2}-E_{J}\cos\left(  \varphi_{n+1} -\varphi_{n}-\frac{2\pi
sB}{\Phi_{0}}y\right)  \right]  . \label{Energy}
\end{equation}
\end{widetext}
where $J$ is the in-plane phase stiffness, $E_{J}$ is the
Josephson energy per unit area, and $w$ is the stack size in the
field direction. To facilitate analysis, we introduce
dimensionless parameters
\[
u=y/\lambda_{J},\ \tilde{L}=L/\lambda_{J},\
e_{J}=\frac{\mathcal{E}_{J} }{E_{J}\lambda_{J} w},\
h=\frac{2\pi\gamma s^{2}B}{\Phi_{0}}.
\]
The reduced energy $e_{J}$ is given by
\begin{equation}
\!e_{J}\!=\!\frac{1}{N}\sum_{n}\!\int\limits_{0}^{\tilde{L}}\!\!du\!\left[
\frac{1}{2}\left( \frac{d\varphi_{n}}{du}\right)
^{2}\!\!-\cos\!\left( \varphi_{n+1}\!-\varphi _{n}\!-\!hu\right)
\right]\!. \label{RedEner}
\end{equation}
In a stable state the phase obeys the following equation, which
expresses the current conservation,
\[
\frac{d^{2}\varphi_{n}}{du^{2}}\!+\!\sin\left(
\varphi_{n+1}\!-\varphi _{n}\!-hu\right)  -\sin\left(
\varphi_{n}\!-\varphi_{n-1}\!-hu\right) \! =\!0
\]
with the boundary conditions (vanishing of the in-plane current at the
boundaries)
\begin{equation}
\left.  \frac{d\varphi_{n}}{du}\right\vert _{u=0,L}=0. \label{BoundCond}
\end{equation}
In principle, a stable state can support a finite Josephson current flowing
through the stack
\begin{equation}
j=j_{J}\lambda_{J}w\int_{0}^{\tilde{L}}du\sin\left(
\varphi_{n+1}-\varphi _{n}-hu\right)  . \label{JosCurrent}
\end{equation}
In a large-size sample this current is concentrated near the
edges. In the following section we will derive a simple closed
expression for the edge current.

\section{\label{Sec:LatStruc}Lattice structure near boundaries. Surface energy and surface current.}

Far from the boundaries (exact criterion will be established
below) the ground state configuration corresponds to the rhombic
lattice (Fig.\ \ref{Fig-DenseLatPhDiag}) and the phase
distribution can be easily calculated using expansion with respect
to the Josephson current \cite{BulClemPRB91,KorshLarkPRB92}
\begin{equation}
\varphi_{n}\approx n\alpha+\frac{n(n+1)}{2}\pi-\frac{2}{h^{2}}\sin\left(
hu-\alpha-\pi n\right)  \label{BulkPhase}
\end{equation}
The bulk part of the energy is degenerate with respect to the
phase shift $\alpha$. Change of $\alpha$ corresponds to
translational displacement of the lattice. This degeneracy is
eliminated by the interaction with the boundaries. To study this
interaction, we introduce the new phase variable $\phi_{n}(u)$
\[
\varphi_{n}=n\alpha+\pi\frac{n(n+1)}{2}+\phi_{n}
\]
and impose the condition that $\phi_{n}$ contains only oscillating
contribution far away from the boundaries, $\phi_{n}(u)\rightarrow
-(-1)^{n}(2/h^{2})\sin\left(  hu-\alpha\right)  $. This new phase $\phi
_{n}(u)$ obeys the following equation
\begin{widetext}
\begin{equation}
\frac{d^{2}\phi_{n}}{du^{2}}+\sin\left(
\phi_{n+1}-\phi_{n}-hu+\alpha+\pi n\right)  +\sin\left(
\phi_{n}-\phi_{n-1}-hu+\alpha+\pi n\right)  =0. \label{PhaseEq}
\end{equation}
\end{widetext}

We assume that the boundary does not change the alternating nature of the
vortex lattice and will look for solution of this equation in the form
\[
\phi_{n}(u)=(-1)^{n}\phi(u).
\]
For the phase $\phi(u)$ we obtain the following equation \cite{KorshLarkPRB92}
\begin{equation}
\frac{d^{2}\phi}{du^{2}}+2\cos2\phi\sin\left(  -hu+\alpha\right)  =0
\label{EqAltPhase}
\end{equation}
with the boundary condition $d\phi/du|_{u=0,L}=0$. The energy and total
Josephson current in terms of $\phi(u)$ are given by
\begin{align}
e_{J}(\alpha)  &  \!=\!\int_{0}^{\tilde{L}}\!du\left[\!
\frac{1}{2}\left( \frac{d\phi}{du}\right)  ^{2}\!-\sin\left(
2\phi\right)  \sin\left(
-hu+\alpha\right)  \!\right]  ,\label{EnAlpha}\\
j(\alpha)  &
=-j_{J}\lambda_{J}\int_{0}^{\tilde{L}}du\sin(2\phi)\cos\left(
-hu+\alpha\right) \nonumber\\
&=j_{J}\lambda_{J}\frac{\partial
e_{J}}{\partial\alpha} \label{CurrAlpha}
\end{align}
Note that the energy and current have the symmetry
$\alpha\rightarrow\alpha +\pi$, $\phi(u)\rightarrow-\phi(u)$
(change of $\alpha$ by $\pi$ is equivalent to vertical
displacement of the lattice by one junction). Therefore, the
energy is $\pi$-periodic function of $\alpha$,
$e_{J}(\alpha+\pi)=e_{J} (\alpha)$.

We now focus on the lattice structure near the boundary $u=0$. In
the limit of high field $h\gg1$ one can separate and average out
the rapidly oscillating part of the phase. This technique has been
used in Ref.\ \onlinecite{KorshLarkPRB92} to study melting of the
Josephson vortex lattice. We split $\phi$ into the smooth ($v$)
and rapidly changing ($\tilde{\phi}$) components,
\[
\phi=v+\tilde{\phi},
\]
where the smooth component describes deformation of the lattice
induced by the boundary and satisfies conditions $dv/du\ll v$ and
$v\rightarrow0$, at large\ $u$. The local lattice compression is
given by $(\partial v/\partial u)/h$.  The oscillating part by
definition obeys the following equation
\begin{equation}
\frac{d^{2}\tilde{\phi}}{du^{2}}+2\cos\left(  2v\right)  \sin\left(
-hu+\alpha\right)  =0. \label{EqOscPhase}
\end{equation}
Neglecting a weak coordinate dependence of $v$ in comparison with
the rapidly oscillating sine function, we obtain an approximate
solution of the last equation
\begin{equation}
\tilde{\phi}\approx\frac{2}{h^{2}}\cos\left(  2v\right)  \sin\left(
-hu+\alpha\right)  \label{OscPhase}
\end{equation}
In the limit $h\gg1$ the oscillating phase $\tilde{\phi}$ can be treated as a
small perturbation. Subtracting Eq. (\ref{EqOscPhase}) from
Eq.\ (\ref{EqAltPhase}) we obtain equation for $v$ in the first order with
respect to $\tilde{\phi}$
\[
\frac{d^{2}v}{du^{2}}-4\tilde{\phi}\sin\left(  2v\right)  \sin\left(
-hu+\alpha\right)  =0
\]
Substituting expression for $\tilde{\phi}$ from Eq.\
(\ref{OscPhase}) and averaging out rapid oscillations, we finally
obtain Sine-Gordon equation for the smooth lattice deformation
$v(u)$,
\begin{equation}
\frac{h^{2}}{2}\frac{d^{2}v}{du^{2}}-\sin\left(  4v\right)  =0.
\label{EqSmooth}
\end{equation}
Using Eq. (\ref{OscPhase}), we also obtain the boundary condition for $v(u)$,
\begin{equation}
\left.  \frac{dv}{du}\right\vert _{u=0}= \left.
-\frac{d\tilde{\phi}}{du}\right\vert _{u=0}=
\frac{2\cos2v_{0}\cos\alpha}{h} \label{BounCondSm}
\end{equation}
with $v_{0}\equiv v(0)$. As follows from Eq.\ (\ref{EqSmooth}) the
boundary deformation decays at the typical length
$l_{h}=h/2\sqrt{2}$ or, in the real units,
\begin{equation}
l_{B}=\gamma s\frac{\pi\gamma s^{2}B}{\sqrt{2}\Phi_{0}}.
\label{Lenght}
\end{equation}
This important length scale appears as an interplay between the
shear and compression stiffnesses of the lattice. A finite system
is in the large-size limit if $L> 2l_{B}$. This condition always
brakes at sufficiently large field, $B>B_{L}$,
\begin{equation}
B_{L}=\frac{L}{\gamma s}\frac{\Phi_{0}}{2\pi\gamma s^{2}}.
\label{BL}
\end{equation}
Averaging out the rapid oscillations, we also derive the surface
energy $e_{s}(\alpha)$ and the surface current $j_{s}(\alpha)$ (in
units of $j_{J} \lambda_{J} w$) in terms of $v(u)$
\begin{align*}
e_{s}(\alpha)  &  \approx\frac{1}{h}\sin\left(  2v_{0}\right)
\cos\left( \alpha\right)  \\
&+\int_{0}^{\infty}du\left[ \frac{1}{2}\left(  \frac{dv}
{du}\right)  ^{2}+\frac{1-\cos4v}{2h^{2}}\right] \\
j_{s}(\alpha)  &  =-\frac{1}{h}\sin\left(  2v_{0}\right)  \sin\left(
\alpha\right)
\end{align*}

Equation (\ref{EqSmooth}) has the well-known soliton solution
\begin{equation}
\tan v=\tan v_{0}\exp\left(  -2\sqrt{2}u/h\right)  \label{vSolution}
\end{equation}
Employing the boundary condition (\ref{BounCondSm}), we derive relation
between the boundary deformation $v_{0}$ and the bulk phase shift $\alpha$
\begin{equation}
\tan\left(  2v_{0}\right)  =-\sqrt{2}\cos\alpha\label{v0alpha}
\end{equation}
From this equation we can see that the maximum deformation
$v_0=\mp\arctan (\sqrt{2})/2\approx \mp 0.478$ occurs at
$\alpha=0,\pi$ and the deformation vanishes at $\alpha=\pi/2$.
\begin{figure}
[ptb]
\includegraphics[width=3.45in ]{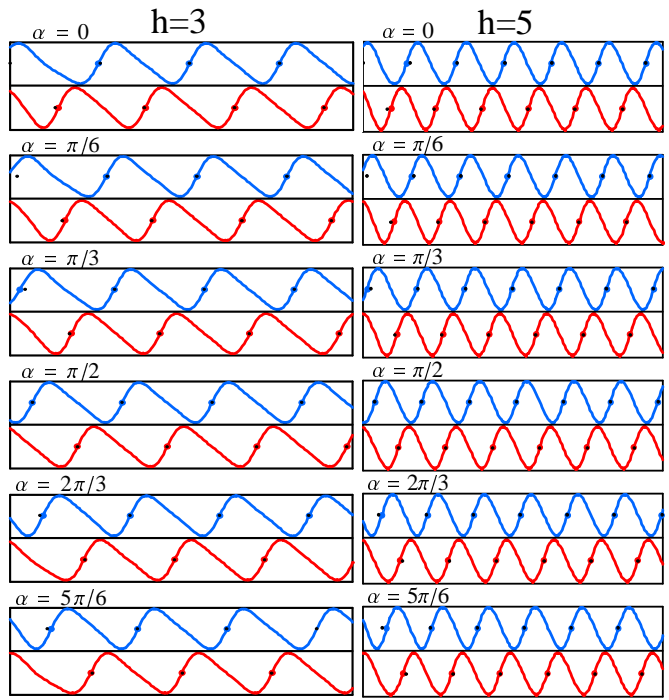}
\caption{Distribution of the Josephson current near the boundary
$u=0$ in the two neighboring junctions for two magnetic fields,
$h=3$ and $5$, and different values of the bulk phase shift
$\alpha$. Small black points mark the positions of the vortex
centers extrapolated from the bulk region assuming undeformed
lattice. The real vortex positions deviate from these marks due to
the surface deformation of the lattice.} \label{Fig-JosCurrSurf}
\end{figure}
Equations (\ref{vSolution}) and (\ref{v0alpha}) determine the
surface deformation $v(u)$ for the given bulk phase shift $\alpha$
(i.e., lattice position). From the surface deformation we can
restore all other surface properties of the lattice. Fig.\
\ref{Fig-JosCurrSurf} shows distribution of the Josephson current
near the boundary in the two neighboring junctions for two
magnetic fields, $h=3$ and $5$, and different values of the
$\alpha$. To illustrates the surface deformation, we also show the
vortex center positions extrapolated from the bulk region assuming
undeformed lattice.

\begin{figure}
[ptb]
\begin{center}
\includegraphics[width=3.45in ]{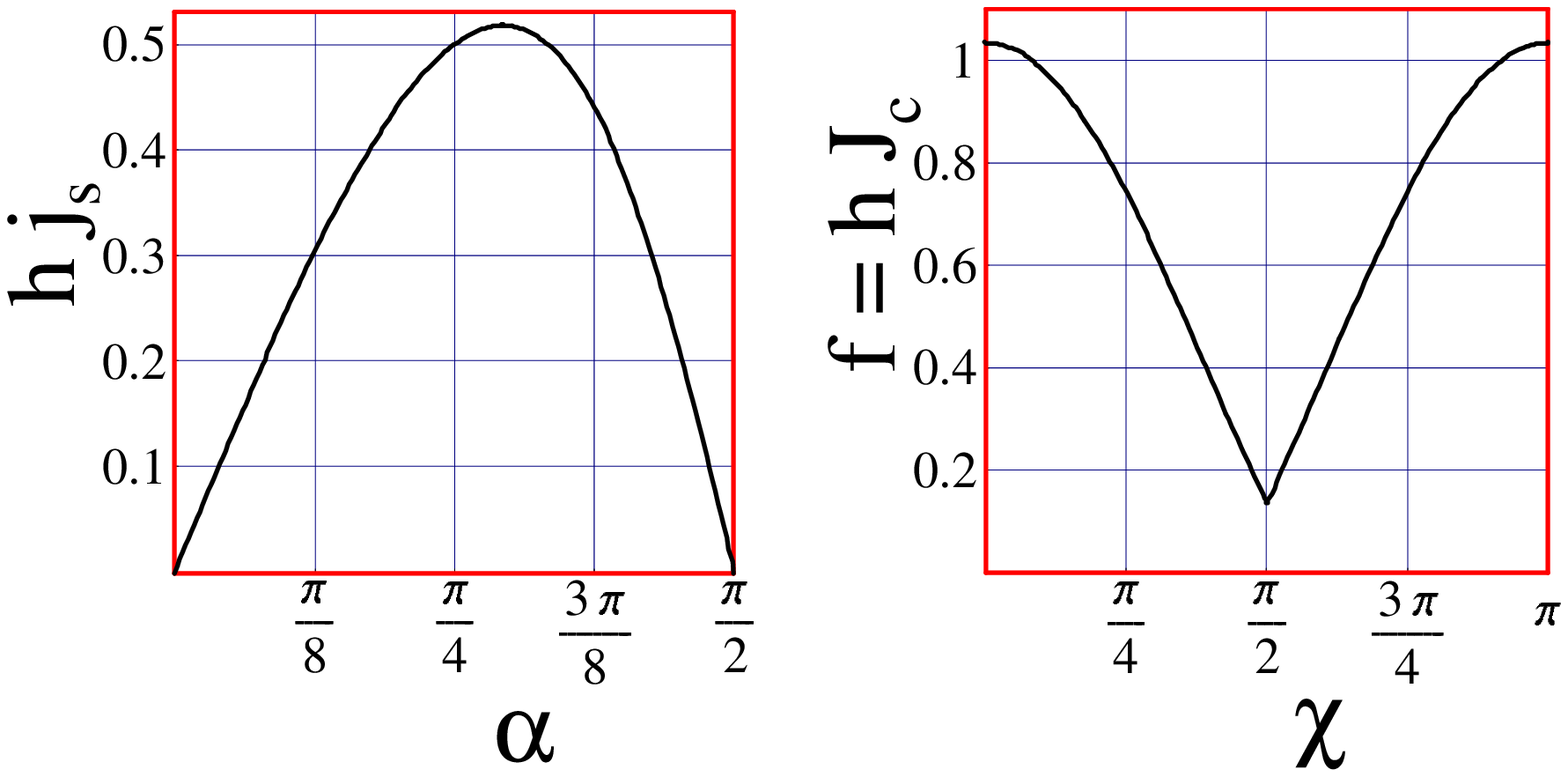} \caption{\emph{Left Panel:} Plot of
the function $hj_s=\frac{1}{\sqrt{2}}\frac{\sin2\alpha
}{\sqrt{2+\cos2\alpha}}$, which gives dependence of the surface
current on the phase shift $\alpha$ (see Eq.\ (\ref{SurfCur})).
\emph{Right Panel:} Dependence of function $f=hJ_{c}$, which
determines the shape of the critical current oscillations in stack
of long junctions,  on the reduced flux through one junction
$\chi=2\pi\Phi/\Phi_{0}$ in Eq.\ (\ref{I_B}). } \label{Fig-hj}
\end{center}
\end{figure}

From the obtained solution for $v(u)$ we derive the closed
analytical expressions for the surface energy and the surface
Josephson current
\begin{align}
e_{s}(\alpha)  &  =\frac{1}{\sqrt{2}h}\left(  1-\sqrt{2+\cos2\alpha}\right)
\label{SurfEn}\\
j_{s}(\alpha)  &  =\frac{1}{\sqrt{2}h}\frac{\sin2\alpha}{\sqrt{2+\cos2\alpha}}
\label{SurfCur}
\end{align}
The $\alpha$-dependence of the surface current is plotted in the
left panel of Fig.\ \ref{Fig-hj}. The maximum surface current,
$j_{s,{\rm max}}$,  is achieved at $\cos 2\alpha=-2+\sqrt{3}$ and
in real units is given by
\begin{equation}
j_{s,{\rm max}}=\sqrt{2-\sqrt{3}}\ j_J\frac{\Phi_0}{2\pi s B}.
\label{MaxEdgeCurr}
\end{equation}
Two possible directions of this current correspond to the Lorentz
force on Josephson vortices directed from the boundary and towards
the boundary, promoting either entrance or exit of vortices. In
contrast to the regime of dilute lattice at $B<B_{cr}$, in which
the maximum values of the entrance and exit current are very
different (see, e.g., Ref.\ \onlinecite{BurPRB96}), in the dense
lattice they are approximately the same. It is interesting to note
that, up to numerical factor, the expression for the entrance
current in the dilute lattice regime coincides with Eq.\
(\ref{MaxEdgeCurr}).

\section{\label{Sec:CritCur}Critical current of a finite stack}

\begin{figure}
[ptbptb]
\begin{center}
\includegraphics[clip,width=3.2in ] {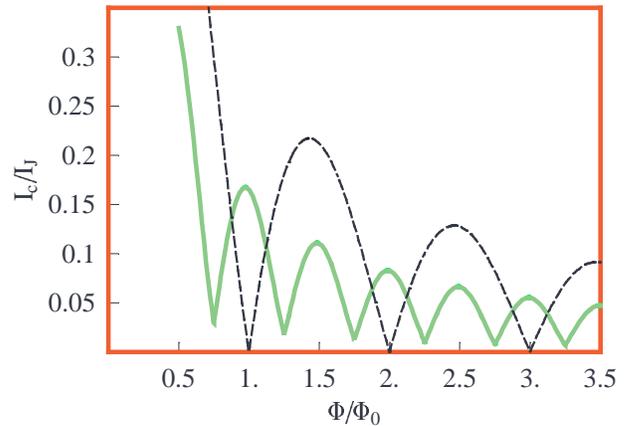} \caption{Thick solid line shows dependence
of the critical current for the stack of long Josephson junctions
on the magnetic flux per junction. For comparison we also show by
dashed line the Fraunhofer dependence $\left\vert \sin(\pi
\Phi/\Phi_{0})\right\vert /(\pi\Phi/\Phi_{0})$ which describes the
field dependence of the critical current in a single small
Josephson junctions.} \label{Fig-Ic-Phi}
\end{center}
\end{figure}

Consider now the total current flowing through a finite stack. For
the given phase shift $\alpha$ (lattice position) the total
current, $J_{L}(\alpha)\equiv J_{L}(\alpha,hL,h)$, is the sum of
the two surface contributions coming from the boundaries $u=0$ and
$u=L$,
\begin{equation}
J_{L}(\alpha)\!=\!\frac{1}{\sqrt{2}h}
\!\left(
\frac{\sin2\alpha}{\sqrt{2\!+\!\cos2\alpha}}
\!-\!\frac{\sin2(hL-\alpha)}{\sqrt{2\!+\!\cos2(hL\!-\!\alpha)}}
\right)
\label{TotalCur}
\end{equation}
Introducing notation $\chi=hL=2\pi\Phi/\Phi_{0}$, where $\Phi=BsL$
is the magnetic flux through one junction, we obtain that the
maximum current $J_{c}(\chi,h)=f(\chi)/h$ is given by
\begin{equation}
J_{c}(\chi,h)=\max_{\alpha}\left[ J_{L}(\alpha,\chi,h)\right]
\label{MAXTotalCur}
\end{equation}
Numerically obtained dependence of $f=hJ_{c}$ vs $\chi$ is plotted
in the right panel of Fig.\ \ref{Fig-hj}. In the real units, the
field dependence of the maximum Josephson current in the field
range $\Phi_{0}/2\pi\gamma s^{2}<B<L\Phi_{0}/\pi\gamma ^{2}s^{3}$
is given by
\begin{equation}
I_{c}(B)=I_{J}\frac{\Phi_{0}}{2\pi sLB}f\left(  \frac{2\pi
sLB}{\Phi_{0} }\right)  , \label{I_B}
\end{equation}
where $I_{J}=j_JLw$ is the maximum Josephson current through the
stack at zero field, and oscillating function $f(\chi)$ is plotted
in the right panel of Fig.\ \ref{Fig-hj}.  To facilitate
comparison with experiment we also obtain an interpolation formula
for $f(\chi)$ for $0<\chi<\pi/2$,
$f(\chi)\approx0.128+0.888\,\cos(\chi)+0.021\,\cos(3\,\chi)$. The
derived field dependence of the critical current somewhat
resembles the well-known Fraunhofer dependence in a single
small-size Josephson junction: it has overall $1/B$ dependence
with large oscillations. However, it has a very different physical
origin and also has several qualitative differences. The most
important difference is that the period of oscillations
corresponds to adding \emph{one flux quantum per two junctions},
i.e., it is two times smaller than for the Fraunhofer
oscillations. This kind of oscillations have been recently
observed in the flux-flow voltage of BSCCO mesas \cite{Ooi01}.
Also, contrary to the Fraunhofer dependence, the points where the
magnetic flux inside the junction equals to integer flux quanta,
correspond to the local maxima of the critical current. Dependence
of the critical current on the magnetic flux per junction is
plotted in Fig.\ \ref{Fig-Ic-Phi}. For comparison, the Fraunhofer
dependence is also shown. The dependence (\ref{I_B}) holds until
two edges give independent contributions to the total current.
This condition breaks when the magnetic field exceeds
$B_{L}=L\Phi_{0} /(2\pi\gamma^{2}s^{3})$. At higher fields the
field dependence crosses over to the usual Fraunhofer dependence.
Recently, this crossover has been studied by numerical simulations
\cite{Machida02}.

\section{\label{Sec:Voltage}Oscillating flux-flow voltage}

\begin{figure}
[ptb]
\begin{center}
\includegraphics[clip,width=3.in ]{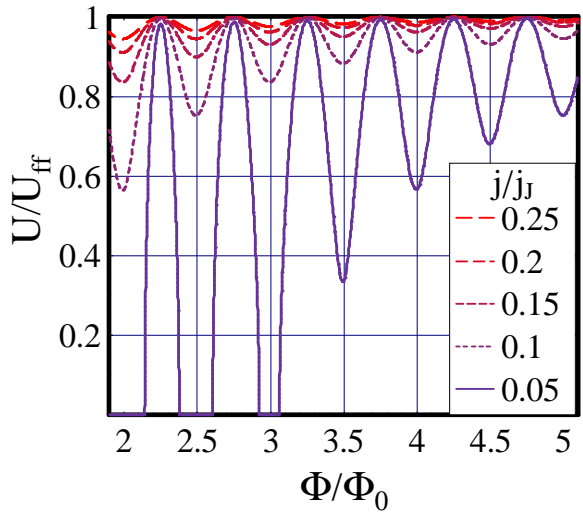} \caption{Oscillating dependencies
of the reduced voltage on the magnetic flux per one junction for
different current densities. The plot represents a universal
behavior for junction stacks with different sizes.}
\label{Fig-UOsc}
\end{center}
\end{figure}

When the external current exceeds the critical current, the
lattice starts to move. At slow motion the surface deformation has
time to adjust to the current lattice position.  In this case the
surface energy produces a periodic potential for the moving
lattice and one can use static results to predict the I-V
dependencies. Time variation of the lattice phase shift obeys
equation
\begin{equation}
\nu_{ff}\frac{d\alpha}{dt}+J_{L}(\alpha)=J,\label{TimeEq}
\end{equation}
where the total surface current $J_{L}(\alpha)\equiv
J_{L}(\alpha,hL,h)$ is given by Eq.\ (\ref{TotalCur}) (for brevity
we again skip in equations its dependence on the magnetic field
and size) and the viscosity coefficient, $\nu_{ff}$, is related to
the flux-flow resistance of the stack, $R_{ff}$,
\[
\nu_{ff}=\frac{N\Phi_{0}}{2\pi cR_{ff}}.
\]
where $N$ is the number of junctions in the stack. Voltage drop
per one junction $U$ is related to $d\alpha/dt$ by the Josephson
relation
\[
U=\frac{\Phi_{0}}{2\pi c}\frac{d\alpha}{dt}.
\]
Solution of Eq. (\ref{TimeEq}) is given by the implicit relation
\[
\int_{0}^{\alpha}\frac{\nu_{ff}d\alpha^{\prime}}{J-J_{L}(\alpha^{\prime})}=t,
\]
from which we obtain the average phase change rate
\[
\overline{\frac{d\alpha}{dt}}=\left[  \frac{1}{\pi}\int_{0}^{\pi}\frac
{\nu_{ff}d\alpha}{J-J_{L}(\alpha)}\right]  ^{-1}
\]
and the flux-flow voltage
\begin{equation}
\frac{U}{U_{ff}}=\left[  \frac{J}{\pi}\int_{0}^{\pi}\frac{d\alpha}
{J-J_{L}(\alpha)}\right]  ^{-1}\label{FluxFlowU}
\end{equation}
with $U_{ff}=R_{ff}J$ being the bare flux-flow voltage without
periodic potential. Because the surface current $J_L(\alpha)\equiv
J_L(\alpha,hL,h)$ oscillates with the magnetic field, this
flux-flow voltage will also experience similar field oscillations.
Such oscillations have been recently observed by Ooi \emph{et.\
al.} \cite{Ooi01} Using scaling property of the current
$J_L(\alpha,hL,h)=L F(\alpha,hL)$ we can see from Eq.\
(\ref{FluxFlowU}) that the reduced flux-flow voltage $U/U_{ff}$
depends only on two parameters: the current density normalized to
the Josephson current density, $i \equiv J/L \equiv j/j_J$, and
the magnetic flux through one junction, $\Phi=BsL$. This allows
for a unified description of the voltage oscillations in junction
stacks with different sizes.  Oscillating dependencies of
$U/U_{ff}$ on the magnetic flux $\Phi$ for different current
densities are shown in Fig.\ \ref{Fig-UOsc}.

At large currents one can expand this result with respect to
surface current and obtain a small correction to the flux-flow
voltage
\begin{equation}
\frac{U}{U_{ff}}\approx1-\frac{g(0)+g(\chi)}{\pi J^{2}h^{2}}
\label{UffLargeJ}
\end{equation}
where the oscillating part is described by dimensionless function
\begin{align*}
g(\chi) &  =\int_{0}^{\pi}\frac{\sin2\alpha\sin2(\alpha-\chi)}{\sqrt
{2+\cos2\alpha}\sqrt{2+\cos2(\chi-\alpha)}}\frac{d\alpha}{\pi},\\
&\approx 0.263 \cos 2\chi +0.0046 \cos 4\chi.
\end{align*}
Therefore, the absolute amplitude of voltage oscillations is given
by
\begin{equation}
\delta U=0.527\frac{R_{ff}}{J}\left(  j_{J}\lambda_{J}\right)
^{2}\left( \frac{\Phi_{0}}{2\pi s\lambda_{J}B}\right)  ^{2}.
\label{AmpUOsc}
\end{equation}
Independently from the dissipation mechanism, the relative
amplitude of oscillations $\delta U/R_{ff}J $ decreases with field
as $1/B^2$. In the regime of dominating in-plane dissipation,
typical for BSCCO, the flux-flow resistance is given by
\cite{JosFluxFlow}
\[
R_{ff}=R_{c}\frac{B^{2}}{B^{2}+B_{\sigma}^{2}};\
B_{\sigma}=\sqrt{\frac
{\sigma_{ab}}{\sigma_{c}}}\frac{\Phi_{0}}{\sqrt{2}\pi\gamma^{2}s^{2}}
\]
where $\sigma_{ab}$ and $\sigma_{c}$ are the components of
quasiparticle conductivity. Using this relation one can rewrite
Eq.\ (\ref{AmpUOsc}) as
\begin{equation}
\delta
U=\frac{R_{ab}J_{M}^{2}}{2J}\frac{1}{1+B^{2}/B_{\sigma}^{2}}
\label{dUff}
\end{equation}
with $J_{M}=c\Phi_{0}/8\pi^{2}\lambda_{ab}^{2}$,
$R_{ab}=Ns/S\sigma_{ab}$, and $S=wL$ is the junction area. For
BSCCO, typically, $B_{\sigma}\sim 10$T. In the region
$B<B_{\sigma}$ the oscillation amplitude weakly depends on the
magnetic field, in agreement with experiment
(see Fig.\ 2 in Ref.\ \onlinecite{Ooi01}). It interesting to note
that
in this regime it is mainly determined
by the in-plane parameters of superconductor.

\section{Summary and acknowledgements}

In conclusions, we found that the edge current of the dense
Josephson lattice is uniquely determined by the magnetic field and
position of the lattice in the bulk. Near the surface the lattice
has alternating deformation, which decays inside superconductor at
the typical length which is proportional to the magnetic field.
Due to the rhombic lattice structure, both the critical current
and the flux-flow voltage at small velocities have oscillating
field dependencies with the period of one flux quantum per two
junctions up to the size-dependent magnetic field.

The author thanks K.\ Hirata and M.\ Machida for useful
discussions. This work was supported by the U.\ S.\ DOE, Office of
Science, under contract \# W-31-109-ENG-38.

\end{document}